\documentclass[a4paper,11pt]{article}

\usepackage{jcappub}

\usepackage{slashed}
\usepackage{bm}
\usepackage{latexsym,amssymb,amsmath,float,url,mathrsfs}
\usepackage{latexsym}
\usepackage{epstopdf}
\usepackage{amsfonts}
\usepackage{amsmath}
\usepackage{amssymb}
\usepackage{comment}
\usepackage{subfigure}
\usepackage{natbib}
\usepackage{hyperref}
\usepackage{pifont}
\usepackage{blindtext}
\usepackage{adjustbox}
\usepackage{multirow}
\usepackage{tabularx}
\usepackage[table]{xcolor}
\usepackage{orcidlink}
\usepackage{tikz}
\usetikzlibrary{tikzmark}
\usepackage{amssymb}
\usepackage{bbold}
\usepackage{blkarray}
\usepackage{graphicx}
\usepackage{amsfonts}
\usepackage{soul}
\usepackage{amssymb}
\usepackage{amsmath}
\usepackage{cancel}
\usepackage{tcolorbox}

\def\xipbh{\xi_\text{\tiny PBH}}

\def\xipbho{\overline{\xi}_\text{\tiny PBH}}

\def\npbh{\overline{n}_\text{\tiny PBH}}

\def\dpbh{\delta_\text{\tiny PBH}}

\def\vx{{\vec{x}}}

\usepackage{graphics,appendix,afterpage,makecell} 

\def\ltap{\ \raisebox{-.4ex}{\rlap{$\sim$}} \raisebox{.4ex}{$<$}\ }

\definecolor{oucrimsonred}{rgb}{0.6, 0.0, 0.0}
\definecolor{persianblue}{rgb}{0.11, 0.22, 0.73}
\definecolor{forestgreen}{rgb}{0.13,0.35,0.13}
\definecolor{lightgray}{rgb}{0.83, 0.83, 0.83}
 \hypersetup{colorlinks, citecolor=oucrimsonred, linkcolor=black, urlcolor=oucrimsonred}
\definecolor{cornellred}{rgb}{0.7, 0.11, 0.11}
\definecolor{navyblue}{rgb}{0.0, 0.0, 0.5}
\definecolor{amethyst}{rgb}{0.6, 0.4, 0.8}
\definecolor{yellow}{rgb}{1.0, 1.0, 0.0}
\definecolor{firebrick}{rgb}{0.7, 0.13, 0.13}
\definecolor{tangerineyellow}{rgb}{1.0, 0.8, 0.0}
\definecolor{deepfuchsia}{rgb}{0.76, 0.33, 0.76}
\definecolor{amber}{rgb}{1.0, 0.75, 0.0}
\definecolor{VioletRed4}{rgb}{0.55, 0.13, .32}
\definecolor{indiagreen}{rgb}{0.07, 0.53, 0.03}
\definecolor{VioletRed4}{rgb}{0.55, 0.13, .32}
\newcommand{\be}{\begin{equation}}
\newcommand{\ee}{\end{equation}}
\newcommand{\bea}{\begin{equation} \begin{aligned}}
\newcommand{\eea}{\end{aligned} \end{equation}}

\definecolor{oucrimsonred}{rgb}{0.6, 0.0, 0.0}
\newcommand\vertarrowbox[3][6ex]{%
  \begin{array}[t]{@{}c@{}} #2 \\
  \left\uparrow\vcenter{\hrule height #1}\right.\kern-\nulldelimiterspace\\
  \makebox[0pt]{\scriptsize#3}
  \end{array}%
}
\hypersetup{
     colorlinks   = true,
     citecolor    = violet,
     urlcolor     = violet,
     linkcolor    = violet}

\definecolor{verdechiaro}{rgb}{0.6,1,0.6}
\definecolor{giallochiaro}{rgb}{1,1,0.6}
\definecolor{bluscuro}{rgb}{0.15, 0.2, 0.9}
\definecolor{verdes}{rgb}{0.1, 0.5, 0.1}%
\definecolor{tangerineyellow}{rgb}{1.0, 0.8, 0.0}

\definecolor{americanrose}{rgb}{1.0, 0.01, 0.24}
\definecolor{cobalt}{rgb}{0.0, 0.28, 0.67}
\definecolor{brandeisblue}{rgb}{0.0, 0.44, 1.0}
\definecolor{mycolor}{rgb}{0.0, 0.0, 0.5}
\definecolor{oxfordblue}{rgb}{0.0, 0.13, 0.28}
\definecolor{azure}{rgb}{0.0, 0.5, 1.0}
\definecolor{turquoiseblue}{rgb}{0.0, 1.0, 0.94}
\newtcolorbox{mynewbox}[1]{colback=white!5!white,colframe=azure!75!black,fonttitle=\bfseries,title=#1}
\newtcolorbox{mybox}{colback=mycolor!5!white,colframe=azure!75!black}
\newtcolorbox{mynamedbox}[1]{colback=mycolor!5!white,colframe=azure!75!black,title=#1}
\definecolor{venetianred}{rgb}{0.78, 0.03, 0.08}
\newtcolorbox{mynamedbox1}[1]{colback=venetianred!5!white,colframe=venetianred!80!black,title=#1}
\newtcolorbox{mynamedbox2}[1]{colback=azure!5!white,colframe=azure!80!black,title=#1}

\newcommand{\td}{{\rm d}}
\newcommand{\Msun}{M_\odot}
\newcommand{\MPBH}{M_{{\text{\tiny PBH}}}}

\definecolor{verdes}{rgb}{0.1, 0.5, 0.1}%
\definecolor{cornellred}{rgb}{0.7, 0.11, 0.11}

\definecolor{VioletRed4}{rgb}{0.55, 0.13, .32}

\definecolor{rossocorsa}{rgb}{0.83, 0.0, 0.0}

\title{The Irrelevance of Primordial Black Hole Clustering in the  LVK    mass range}
\author[a,b]{F.~Crescimbeni,}
\author[c]{V.~Desjacques,} 
\author[d]{G.~Franciolini,}
\author[e]{A.~Ianniccari,}
\author[*,e]{A.J.~Iovino,}
\author[f,g]{G.~Perna,}
\author[e]{D.~Perrone,}
\author[e]{A.~Riotto,}
\author[h]{H.~Veerm\"ae}

\affiliation[a]{Dipartimento di Fisica, ``Sapienza'' Universit\`a di Roma,
\\Piazzale Aldo Moro 5, 00185, Roma, Italy}
\affiliation[b]{Istituto Nazionale di Fisica Nucleare, sezione di Roma,
\\Piazzale Aldo Moro 5, 00185, Roma, Italy}
\affiliation[c]{Physics department, Technion, 3200003 Haifa, Israel}
\affiliation[d]{CERN, Theoretical Physics Department, 
\\Esplanade des Particules 1, Geneva 1211, Switzerland}
\affiliation[e]{Department of Theoretical Physics and Gravitational Wave Science Center,  \\
24 quai E. Ansermet, CH-1211 Geneva 4, Switzerland}
\affiliation[f]{Dipartimento di Fisica e Astronomia ``Galileo Galilei'', Universit\`a degli Studi di Padova, Via Marzolo 8, I-35131, Padova, Italy}
\affiliation[g]{INFN, Sezione di Padova, Via Marzolo 8, I-35131, Padova, Italy}
\affiliation[h]{Keemilise ja Bioloogilise F\"u\"usika Instituut, R\"avala pst. 10, 10143 Tallinn, Estonia}

\emailAdd{ antoniojunior.iovino@uniroma1.it}

\rightline{CERN-TH-2025-027}

\abstract{We show that in common inflationary models where primordial black holes are formed due to the collapse of sizeable inflationary perturbations, their initial spatial clustering beyond the Poisson distribution does not affect the binary mergers— including sub-solar primordial black holes—responsible for the gravitational waves detectable by LIGO-Virgo-KAGRA. This is a consequence of the existing FIRAS CMB distortion constraints on the relevant scales. However, this conclusion may not hold for lighter masses potentially accessible by future gravitational wave observations and for multi-field inflation scenarios, where the curvature perturbation can acquire a large-scale modulation from a field that is not the primary source of the perturbation. }

\begin{document}
\maketitle
\flushbottom
\section{Introduction}
The physics of Primordial Black Holes (PBHs) has garnered significant interest in recent years (see~\cite{Sasaki:2018dmp, Carr:2020gox, Green:2020jor,LISACosmologyWorkingGroup:2023njw,Riotto:2024ayo} for some reviews), largely due to the numerous detections of gravitational waves (GWs) originating from BH binary mergers~\cite{LIGOScientific:2016aoc, LIGOScientific:2018mvr, LIGOScientific:2020ibl, LIGOScientific:2021djp, KAGRA:2021vkt} and the hypothesis that some of these may be of primordial origin~\cite{Bird:2016dcv, Sasaki:2016jop, Clesse:2016vqa,Mukherjee:2021ags,Afroz:2024fzp}.  

Various mechanisms and models have been proposed as viable pathways for PBH production. Depending on the specific formation scenario and the models for PBH formation, a wide range of mass functions have been predicted (see~\cite{LISACosmologyWorkingGroup:2023njw} for a recent review). In this work, we adopt the standard formation scenario, where PBHs arise from the gravitational collapse of large over-densities in the primordial density contrast field upon their horizon re-entry~\cite{Ivanov:1994pa, Ivanov:1997ia, Blinnikov:2016bxu}.  
Thus, to achieve a non-negligible PBH abundance, it is necessary to have an enhancement of the primordial curvature power spectrum on scales smaller than those probed by the Cosmic Microwave Background (CMB), where its spectral amplitude is approximately $10^{-9}$~\cite{Planck:2018jri}. In Ultra-Slow-Roll (USR) models of single-field inflation, for example, the peak in the power spectrum of the curvature perturbation arises from a brief phase of ultra-slow-roll which is typically followed by slow-roll or constant-roll inflation \cite{Ivanov:1994pa, Leach:2000ea, Bugaev:2008gw, Alabidi:2009bk, Drees:2011hb, Drees:2011yz, Alabidi:2012ex, Kannike:2017bxn, Ballesteros:2017fsr, Di:2017ndc, Germani:2017bcs, Cicoli:2018asa, Ozsoy:2018flq, Bhaumik:2019tvl, Ballesteros:2020qam, Karam:2022nym, Franciolini:2022pav, Balaji:2022rsy, Allegrini:2024ooy, Briaud:2025hra}. In curvaton-like models, instead, the enhancement is due to an extra light field whose perturbations contribute to the curvature perturbation at the time of decay \cite{Enqvist:2001zp,Lyth:2001nq,Sloth:2002xn,Lyth:2002my,Dimopoulos:2003ii,Kohri:2012yw,Kawasaki:2012wr,Kawasaki:2013xsa,Carr:2017edp,Ando:2017veq,Ando:2018nge,Chen:2019zza,Liu:2020zzv,Pi:2021dft,Cai:2021wzd,Liu:2021rgq,Chen:2023lou,Cable:2023lca,Gow:2023zzp,Inomata:2023drn,Ferrante:2023bgz}. In both cases, the enhanced perturbations are accompanied by some level of non-Gaussianity (NG) \cite{Atal:2019cdz,Sasaki:2006kq,Pi:2022ysn}.

The presence of NG induces a correlation between small and large scales. Hence PBHs are never exactly Poisson distributed at the time of formation, but are always correlated up to a given scale: small scales where the PBH is formed, in fact, are modulated by the large-scale fluctuations. Of course, the importance of clustering associated to PBH is significant only if such correlation is larger than the Poisson contribution. In addition, the initial clustering could influence the merger rate of PBH binaries, whose GWs could be detected by current and future experiments, with particular interest for the subsolar BHs which might be a smoking gun for their primordial nature \cite{Franciolini:2021xbq,Crescimbeni:2024cwh}. 

We show in this work that current observations prevent such a possibility for PBH masses in the LIGO-Virgo-KAGRA (LVK) mass range. The reasoning is the following. For PBH clustering to meaningfully affect the merger rate of PBH binaries within the (sub)-solar mass range, PBHs must exhibit spatial correlations on scales of comoving kpc relevant to the current merger rate. Even by maximizing clustering with a broad curvature perturbation spectrum, current bounds on CMB distortions~\cite{Fixsen:1996nj,Chluba:2012gq,Chluba:2012we,Chluba:2013dna,Bianchini:2022dqh,Iovino:2024tyg} significantly constrain the range of scales where clustering may play a role, see Ref.\,\cite{DeLuca:2021hcf}. In the remaining relevant range of scales, the models commonly studied in the literature do not deliver a large enough NG to overcome the Poisson distribution. However,  we will also show that, for PBH masses smaller than about $10^{-6}\ M_\odot$, whose binaries could lead to a signal  potentially detectable by future experiments, clustering may indeed play a role. 

The paper is organized as follows. In section 2, we briefly describe how to compute the PBH abundance in the presence of local non-Gaussianity in the curvature perturbation field, following the prescription based on threshold statistics on the compaction function. In section 3, we describe how to compute the correlation length and the bias factor in the presence of initial clustering PBHs, and then we compute these quantities for some realistic scenarios. In section 4, we show how in the relevant scales for the LVK observations and Einstein Telescope the spatial clustering does not play any role in the binary mergers. We conclude in section 5. The paper is also supplemented by one Appendix.

\section{The PBH abundance}
In this section, we summarize the formalism to evaluate the PBH abundance and set some useful definitions. The criterion we adapt for the formation of a PBH in a given region and for estimating the amount of clustering is based on the compaction function~\cite{Shibata:1999zs}. It is defined as twice the local mass $(M)$ excess relative to the background value $(M_b)$, divided by the areal radius $R(r,t)=a(t)\, e^\zeta\, r$ (in terms of the scale factor $a$ and the comoving curvature perturbation $\zeta$)
\begin{align}\label{eq:DefinitionCompaction}
\mathcal{C}(r,t) = \frac{2\left[M(r,t) - M_b(r,t)\right]}{R(r,t)} =
\frac{2}{R(r,t)}\int_{V_{R}} d^{3}\vec{x}\,
\rho_b(t)\delta(\vec{x},t).
\end{align}
On super-horizon scales, adopting the gradient expansion approximation, and assuming spherical symmetry, the density contrast is~\cite{Harada:2015yda}
\begin{align}\label{eq:SphericalDelta}
\delta(r,t) = 
-\frac{4}{9}
\left(\frac{1}{aH}\right)^2 
e^{-2\zeta(r)}\left[
\zeta^{\prime\prime}(r) + \frac{2}{r}\zeta^{\prime}(r) + \frac{1}{2}\zeta^{\prime 2}(r)
\right]\,,
\end{align}
where $' \equiv d/dr$, the factor 4/9 is for a radiation-dominated universe, and $\zeta(r)$ is assumed to be constant on super-horizon scales. We can use the super-horizon expansion since we are interested in the correlation on scales larger than the Hubble radius within which PBHs form.

In substituting the previous expression in Eq. \eqref{eq:DefinitionCompaction} and performing the volume integral, the compaction function takes the form~\cite{Harada:2015yda}
\begin{align}\label{eq:CompactionFull}
\mathcal{C}(r) = 
-\frac{4}{3}\,r\,\zeta^{\prime}(r)\left[
1 + \frac{r}{2}\zeta^{\prime}(r)
\right] = 
\mathcal{C}_1(r) - \frac{3}{8}\mathcal{C}^2_1(r),~~~~~~~~~~~~~
\mathcal{C}_1(r) = -\frac{4}{3} r\zeta^{\prime}(r).
\end{align} 
Notice that $\mathcal{C}$ becomes time-independent and
Eq.\,\eqref{eq:CompactionFull} includes the full non-linear relation between $\delta$ and $\zeta$. 
We define $r_m$ as the scale at which the compaction function is maximized. Therefore, it verifies the condition 
\begin{equation}
\mathcal{C}^{\prime}(r_m) = 0
~~~~~~~~~~\text{that is}~~~~~~~~~~
\zeta^{\prime}(r_m) + r_m\zeta^{\prime\prime}(r_m) = 0    
\end{equation}
in terms of 
the comoving curvature perturbation. 
If we define $\mathcal{C}_{\text{\tiny max}} = \mathcal{C}(r_m)$ as the value of the compaction at the position of the maximum, 
PBHs form only if the maximum value of the compaction function exceeds some threshold value, 
$\mathcal{C}_\text{\tiny max} > \mathcal{C}_{\text{\tiny c}}$. Notice also that, at the horizon crossing of 
the relevant scale $r_m = (aH)^{-1}$, 
the compaction function at its peak becomes equal to the fully non-linear density contrast smoothed over the horizon volume, which is the quantity we will compute the correlation of.

A crucial role in determining the PBH abundance, as well as the clustering, is the NG in the curvature perturbation \,\cite{Young:2013oia,Bugaev:2013vba,Young:2014ana,Nakama:2016gzw,Byrnes:2012yx,Franciolini:2018vbk,Yoo:2018kvb,Kawasaki:2019mbl, Riccardi:2021rlf,Taoso:2021uvl,Biagetti:2021eep,Kitajima:2021fpq,Hooshangi:2021ubn,Meng:2022ixx,Young:2022phe,Escriva:2022pnz,Hooshangi:2023kss,Ianniccari:2024bkh,Ferrante:2022mui}.
In the literature, the local non-Gaussian behavior of the curvature perturbation $\zeta$ is usually parameterized by the expansion\,\cite{Gangui:1993tt,Komatsu:2001rj}
\begin{align}\label{eq:FirstExpansion}
\zeta = \zeta_\text{\tiny g} + \frac{3}{5}f_\text{\tiny NL}\zeta_\text{\tiny g}^2
 + \frac{9}{25}g_\text{\tiny NL}\zeta_\text{\tiny g}^3 + \dots\,,
\end{align}
where $\zeta_{\text{\tiny g}}$ obeys the Gaussian statistics while the parameters $f_\text{\tiny NL}$, $g_\text{\tiny NL}$, $\dots$ (which, in full generality, depending on the scale of the perturbation) encode deviations from the Gaussian limit. 

Generally, when a closed-form resummed expression for $\zeta$ is available, it has been shown that truncating the previous power series expansion at a fixed order can lead to an incorrect estimation of the PBH abundance\,\cite{Ferrante:2022mui,Iovino:2024crh}. Consequently, to maintain model independence in this part of the work, we introduce primordial NGs through the following functional form
\be\label{eq:zeta}
\zeta = F(\zeta_\text{\tiny g}). 
\ee
In a radiation-dominated universe, the linear component of the compaction function takes the form
\begin{align}\label{eq:C1expl}
\mathcal{C}_1(r) = -\frac{4}{3}\,r\,\zeta_\text{\tiny g}^{\prime}(r)\,
\frac{dF}{d\zeta_\text{\tiny g}} = 
\mathcal{C}_\text{\tiny g}(r)\,
\frac{dF}{d\zeta_\text{\tiny g}},~~~{\rm with}~~~
\mathcal{C}_\text{\tiny g}(r) =
-\frac{4}{3}r\zeta_\text{\tiny g}^{\prime}(r).
\end{align}
Consequently, the compaction function reads
 \begin{align}\label{eq:CCgau}
\mathcal{C}(r) = 
\mathcal{C}_\text{\tiny g}(r)\,
F' - \frac{3}{8}
 \mathcal{C}^2_\text{\tiny g}(r)
 \left(F'
 \right)^2,
 \end{align}
where primes of the function $F$ indicate derivatives with respect to $\zeta_\text{\tiny g}$.
The compaction function thus depends on both the Gaussian linear component $\mathcal{C}_\text{\tiny g}$ and the Gaussian curvature perturbation $\zeta_\text{\tiny g}$. 
Both are Gaussian random variables since $\zeta_\text{\tiny g}$ is Gaussian by definition, while $\mathcal{C}_\text{\tiny g}$ is proportional to the derivative $\zeta_\text{\tiny g}'$. 
We write~\cite{Young:2022phe}
\be
\mathcal{C}_\text{\tiny g}(r)=-\frac{4}{9}r^2\int \td^3y\,\nabla^2\zeta_\text{\tiny g}(\vec y)\,W(\vec x-\vec y,r)
\ee
and 
\be
\zeta_\text{\tiny g}(r)=\int \td^3y \,\zeta_\text{\tiny g}(\vec y)\,W_{\text{\tiny s}}(\vec x-\vec y,r),
\ee
where $W_{\text{\tiny s}}$ is the spherical-shell window with Fourier transform $W_{\text{\tiny s}}(k,r) = \sin(kr)/kr$ and $W$ is the Heaviside-step function with Fourier transform
\be
W(k,r)   =   3\left[
\frac{\sin(kr) - kr\cos(kr)}{(kr)^3}\right].
\ee
The two-dimensional joint PDF of $\zeta_\text{\tiny g}$ and $\mathcal{C}_\text{\tiny g}$ can be written as
 \begin{align}\label{eq:PDFCompa}
 \textrm{P}_\text{\tiny g}(\mathcal{C}_\text{\tiny g},\zeta_\text{\tiny g}) 
 = \frac{1}{(2\pi)\sqrt{\det\Sigma_1}}
 \exp\left(
 -\frac{1}{2}\vec{Y}_1^{\rm T}\Sigma_1^{-1}\vec{Y}_1
 \right),~~\vec{Y}_1 =\left(
\begin{array}{c}
 \mathcal{C}_\text{\tiny g}  \\
  \zeta_\text{\tiny g}    
\end{array}
\right),~~
\Sigma_1 =
\left(
\begin{array}{cc}
 \sigma_{\text{\tiny c}}^2 &
 \sigma_{\text{\tiny cr}}^2 \\ \sigma_{\text{\tiny cr}}^2 &\sigma_{\text{\tiny r}}^2    
\end{array}
\right),
 \end{align}
 where $\Sigma_1=\langle \vec{Y}_1\vec{Y}_1^{\rm T}\rangle$ is the covariance matrix. The entries of $\Sigma_1$ are
 \begin{equation}
 \sigma_{\text{\tiny c}}^2  =\langle\mathcal{C}_\text{\tiny g}\mathcal{C}_\text{\tiny g}\rangle  = 
  \frac{16}{81}\int_0^{\infty}\frac{\td k}{k}
  (kr_m)^4 W^2(k,r_m)  \!P^{T}_\zeta(k)\!,\label{eq:Var1}
  \end{equation}
   \begin{equation}
 \sigma_{\text{\tiny cr}}^2  =\langle\mathcal{C}_\text{\tiny g}\zeta_\text{\tiny g}\rangle  =  
 \frac{4}{9}\int_0^{\infty}\frac{\td k}{k}(kr_m)^2
 W(k,r_m)
 W_{\text{\tiny s}}(k,r_m)  \!P^{T}_\zeta(k)\! ,
 \end{equation}
  \begin{equation}
  \sigma_{\text{\tiny r}}^2  =\langle\zeta_\text{\tiny g}\zeta_\text{\tiny g}\rangle  =    \int_0^{\infty}\frac{\td k}{k}
  W_{\text{\tiny s}}^2(k,r_m) \!P^{T}_\zeta(k)\!,\label{eq:Var3}
 \end{equation}
with $P^{T}_\zeta=T^2\left(k, r_m\right) P_\zeta(k)$, where $T\left(k, r_m\right)$ is the radiation transfer function and all the entries are evaluated at $r_m=1/aH$.

After computing the inverse of $\Sigma_1$ and its determinant, and completing the square in the argument of the exponential function,
 Eq.\,\eqref{eq:PDFCompa} can be recast in the form
\begin{align}\label{eq:PDFCompa2}
 \textrm{P}_\text{\tiny g}(\mathcal{C}_\text{\tiny g},\zeta_\text{\tiny g}) =
 \frac{1}{(2\pi)\sigma_{\text{\tiny c}}\sigma_{\text{\tiny r}}\sqrt{1-\gamma_{\text{\tiny cr}}^2}}
 \exp\left(
 -\frac{\zeta_\text{\tiny g}^2}{2\sigma_{\text{\tiny r}}^2}
 \right)
 \exp\left[
 -\frac{1}{2(1-\gamma_{\text{\tiny cr}}^2)}\left(
 \frac{\mathcal{C}_\text{\tiny g}}{\sigma_{\text{\tiny c}}} - \frac{\gamma_{\text{\tiny cr}}\zeta_\text{\tiny g}}{\sigma_{\text{\tiny r}}}
 \right)^2
 \right],
\end{align}
where 
\be
\gamma_{\text{\tiny cr}} = \frac{\sigma_{\text{\tiny cr}}^2}{\sigma_{\text{\tiny c}} \sigma_{\text{\tiny r}}}.\label{eq:GammacrDef}
\ee
Using the conservation of the probability, we can therefore write
\begin{eqnarray}
     P\left(\mathcal{C} > \mathcal{C}_{\text{\tiny c}}\right) 
     &=&
    \int_{\mathcal{D}} \textrm{P}_\text{\tiny g}(\mathcal{C}_\text{\tiny g},\zeta_\text{\tiny g})\td\mathcal{C}_\text{\tiny g} \td\zeta_\text{\tiny g}\,\\
 \mathcal{D}  
    &=& \label{eq:Domain}
    \left\{
\mathcal{C}_\text{\tiny g},\,\zeta_\text{\tiny g} \in \mathbb{R}:
\mathcal{C}(\mathcal{C}_\text{\tiny g},\zeta_\text{\tiny g}) > \mathcal{C}_{\text{\tiny c}}
~\land~\mathcal{C}_1(\mathcal{C}_\text{\tiny g},\zeta_\text{\tiny g}) < 4/3
\right\}.\label{eq:RegionD}
\end{eqnarray}
We are finally in the position to give our prescription to calculate the PBH abundance, following Ref.~\cite{Ferrante:2022mui} (see also \cite{Gow:2022jfb}) based on threshold statistics\footnote{A discrepancy is present between peak theory and threshold statistics (see, e.g., Refs.~\cite{Green:2004wb,Young:2014ana,Pi:2024ert}). A technical drawback of peaks theory is that it is not clear how to include NGs in the computation of the abundance using a generic functional form for the curvature perturbation field as in Eq.~\eqref{eq:zeta}. Hence, for comparison, we limit our analysis to the threshold statistic approach.} on the compaction function $\mathcal{C}$.
The abundance of PBHs is given by the integral (see e.g.~\cite{Karam:2022nym})
\bea
    f_{\text{\tiny PBH}}\left(\MPBH\right)
    &\equiv 
    \frac{1}{\Omega_{\rm  DM}}\frac{ \td \Omega_{\text{\tiny PBH}}}{\td \ln \MPBH}  \\
    &=
    \frac{1}{\Omega_{\rm  DM}}
    \int \td \ln M_{H} \left (\frac{M_{H}}{M_\odot}\right )^{-1/2}
    \left(\frac{g_{*s}^4/g_*^3}{106.75} \right)^{-\frac14}\!
    \left (\frac{\beta(M_\text{\tiny PBH},M_\text{\tiny H})}{7.9\times 10^{-10}} \right)\,,
\eea
where $\Omega_{\rm  DM} = 0.264$ is the cold dark matter density of the universe, so the total abundance of PBHs is
\be
    f_{{\text{\tiny PBH}}}=\int f_{{\text{\tiny PBH}}}\left(\MPBH\right) \td \ln \MPBH.
\ee

The relation between the mass of the resulting PBH $M_\text{\tiny PBH}$ and the horizon mass $M_\text{\tiny H}$ is dictated by the following critical scaling law\,\cite{Choptuik:1993,Evans:1994}
\be
    M_\text{\tiny PBH}(\mathcal{C}) = \mathcal{K} M_\text{\tiny H} (\mathcal{C} - \mathcal{C}_{\text{\tiny c}})^{\gamma}\,,
\ee
with $\gamma=0.38$\,\cite{Musco:2008hv,Ianniccari:2024ltb}. The horizon mass corresponds to the $k$ with the relation
\begin{equation}
    M_{\text{\tiny H}} \simeq 17 \Msun\left(\frac{g_{\star}}{10.75}\right)^{-1 / 6}\left(\frac{k / \kappa}{10^6 \mathrm{Mpc}^{-1}}\right)^{-2}\,
\end{equation}
where the factor $\kappa$ depends on the shape of the power spectrum (see Ref.\,\cite{Musco:2020jjb}) and in this work is fixed to $\kappa=4.5$. The mass fraction $\beta$ is obtained from the joint probability distribution function $\rm{P}_{\mathrm{g}}$ 
\begin{equation}\label{eq:beta}
    \beta(M_\text{\tiny PBH},M_{H}) = 
    \int_{\mathcal{D}} \frac{M_\text{\tiny PBH}}{M_{H}}
    \delta\left[ \ln\frac{M_\text{\tiny PBH}}{M_\text{\tiny PBH}(\mathcal{C})} \right] 
    {\rm P}_{g}(\mathcal{C}_{\text{\tiny g}},\zeta_{\text{\tiny g}})
    \td\mathcal{C}_{\text{\tiny g}}\td\zeta_{\text{\tiny g}}\,,
\end{equation}
where the domain of integration is given by Eq.\eqref{eq:RegionD}, the multivariate Gaussian is given in Eq.\,\eqref{eq:PDFCompa2} and the correlators are computed as in Eqs.\,\eqref{eq:Var1}-\eqref{eq:Var3}. In this work, we have followed the prescription given in Ref.\,\cite{Musco:2020jjb} to compute the values of the threshold $\mathcal{C}_{\text{\tiny c}}$ and the position of the maximum of the compaction function $r_m$, which depend on the shape of the power spectrum. We get $\mathcal{C}_{\text{\tiny c}}=0.56$. The presence of the QCD phase transitions is taken into account by considering that $\gamma\left(M_{\text{\tiny H}}\right), \mathcal{K}\left(M_{\text{\tiny H}}\right), \mathcal{C}_{\text{\tiny c}}\left(M_{\text{\tiny H}}\right)$ and $\Phi\left(M_{\text{\tiny H}}\right)$ are functions of the horizon mass around $\MPBH=\mathcal{O}\left(\Msun\right)$\,\cite{Franciolini:2022tfm,Musco:2023dak}. As we will see in the next sections, the latter has an important phenomenological consequence when a PBH mass function centered nearly around one solar mass is considered.

\section{The PBH clustering}

Having described how to estimate the PBH abundance, we now turn our attention to the spatial clustering of PBHs.
To characterize the PBH two-point correlation function $\xipbh(x)$ (or, simply, correlation function) at any comoving separation $x=|\vx|$, we can use 
the overdensity of discrete PBH centers at position ${\vx}_i$ (eventually smoothed over a sphere with a radius equal to the Hubble radius at the moment the perturbations re-enter the horizon)
\begin{equation}\label{eq:Cluster1}
\dpbh({\vec x})=\frac{1}{\npbh}\sum_i \delta_D(\vx-\vx_i)-1 \;,
\end{equation}
where $\delta_D(\vx)$ is the three-dimensional Dirac distribution, $\npbh$ is the average comoving number density of PBH, and $i$ runs over the initial positions of PBH.
The corresponding two-point correlation function must take the general form (see, for instance, Ref. \cite{Baldauf:2013hka} in the context of large
scale structure)
\begin{eqnarray}
\big\langle\dpbh(\vx)\dpbh(0) \big\rangle &=& \frac{1}{\npbh}\delta_D(\vx)-1
+\frac{1}{\npbh^2}\Big<\sum_{i\neq j}\delta_D(\vx-\vx_i)\delta_D(\vx_j)\Big>\nonumber\\
&=&
\frac{1}{\npbh}\delta_D(\vx)+ \xipbh(x) \;,
\label{eq:PBH2pt}
\end{eqnarray}
where\,\cite{DeLuca:2022bjs}
\begin{equation}\label{eq:n_PBH}
\overline{n}_{\text{\tiny PBH}} \simeq 30 f_{\text{\tiny PBH}}\left(\frac{M_\text{\tiny PBH}}{\Msun}\right)^{-1} \mathrm{kpc}^{-3}.
\end{equation}
Here, $\xipbh(x)$ is the reduced PBH correlation function.
We would like to evaluate the correlation function $\xipbh$ between PBHs at a comoving distance $x$. Since PBHs form at peaks of the underlying radiation overdensity, their two-point correlator is biased with respect to the one of radiation \cite{Desjacques:2018wuu}. For separations $x$ much larger than the typical (comoving) size of perturbations collapsing into PBHs, we have
\begin{equation}
    \xipbh (x) \approx b_1^2 \xi_{\text{\tiny r}}(x),
\end{equation}
where $\xi_{\text{\tiny r}}$ is the correlation function for radiation and $b_1$ is the linear bias factor, which will be computed later in this work. The initial PBH two-point function can also be expressed as\,\cite{Bardeen:1985tr}
\begin{equation}\label{eq:xipbh}
    \xipbh (x)=\frac{1}{2\pi^2}\int_0^{\infty} \td k \, k^2 P_{\text{\tiny PBH}}(k) j_0(k x)\,,
\end{equation}
where $j_0(k x)$ is a spherical Bessel function, $a_{\text{\tiny H}}$ is the scale factor evaluated at PBH formation time and 
\begin{equation}
    P_{\text{\tiny PBH}}(k) =\left(\frac{4}{9}\right)^2 b_1^2 P_{\zeta}(k)\,.
    \label{PS_PBH}
    \end{equation}
To be as conservative as possible in our conclusions, we focus on the finite scale invariant curvature power spectrum that maximizes the clustering, i.e. a broad spectrum that correlates small and large scales
\begin{equation}\label{eq:PS}
    P_{\zeta}(k)= A_{\text{\tiny s}}\frac{2\pi^2}{k^3} \theta(k_{\text{\tiny max}}-k)\theta(k-k_{\text{\tiny min}})\,,
\end{equation}
where $k_{\text{\tiny min}}$ and $k_{\text{\tiny max}}=k_{\text{\tiny min}}\cdot \Delta $ define the minimum and maximum scales, with $\Delta $ being the dimensionless width of the spectrum. With such a power spectrum, the lightest PBHs form when the smallest scale $\sim k^{-1}_{\text{\tiny max}}$ re-enters the horizon. They dominate the mass function over the heavier PBHs formed later because their density is diluted much slower than radiation  \cite{MoradinezhadDizgah:2019wjf,DeLuca:2020ioi}. As we will show, such light PBHs are clustered up to the scale $\sim k^{-1}_{\text{\tiny min}}$. The integral in Eq. \eqref{eq:xipbh} can be solved analytically, giving 
\bea\label{eq:xiaH}
    {\xi}_\text{\tiny PBH}(x) 
    &= \left(\frac{4}{9}\right)^2 A_s b_1^2 \left[ 
    \operatorname{Ci}(k_{\text{\tiny max}} x) 
-   \operatorname{Ci}(k_{\text{\tiny min}} x) 
-   \frac{\sin(k_{\text{\tiny max}} x)}{k_{\text{\tiny max}}x} 
+   \frac{\sin(k_{\text{\tiny min}} x)}{k_{\text{\tiny min}}x} \right] \\
    &\approx \left(\frac{4}{9}\right)^2 A_s b_1^2 
    \left\{\begin{array}{ll}
        \ln\Delta\,, & \quad x \ll k^{-1}_{\text{\tiny{ max}}}, \\
        \ln\left(\frac{1.53}{x k_{\text{\tiny{min}}} }\right)\,, & \quad k^{-1}_{\text{\tiny{ max}}} \ll x \ll k^{-1}_{\text{\tiny{ min}}}, \\
        \mathcal{O}((x k_{\text{\tiny{min}}})^{-2})\,, & \quad x \gg k^{-1}_{\text{\tiny{ min}}},
    \end{array}\right.
\eea
where 
$
    {\rm{Ci}}(z) \equiv -\int_z^{\infty} \frac{\cos(t)}{t} {\rm d} t\,
$
is the cosine integral. The second line gives the asymptotics in the regions separated by $k^{-1}_{\text{\tiny{ max}}}$ and $k^{-2}_{\text{\tiny{ max}}}$, which hold well for $\Delta \gg1$, that is, for the broad spectra considered in this study. Thus, $\xipbh (x)$ reaches its maximum at smallest distances $x \lesssim k^{-1}_{\text{\tiny{ max}}}$, where it stays constant. At the largest distances, $x \gtrsim k^{-1}_{\text{\tiny{ min}}}$ is a rapidly damped oscillating function. These oscillations are partly an artefact of the sharp cuts in the adapted power spectrum \eqref{eq:PS}. However, they are not relevant to our analysis, which depends mostly on the intermediate range $k^{-1}_{\text{\tiny{max}}} \lesssim x \lesssim k^{-1}_{\text{\tiny{ min}}}$. We also remark, that the first zero of $\xipbh (x)$ lies at $x = 2.16/k_{\text{\tiny{max}}}$, that is, at slightly larger distances than suggested by the approximation in the intermediate region.

The average number density of PBHs in a volume of radius $R$ is
\be
    \langle N(R)\rangle
    =\overline{n}_{\text{\tiny PBH}} V(R)+\overline{n}_{\text{\tiny PBH}} \int_0^R\td^3 x\, \xipbh(x)=\overline{n}_{\text{\tiny PBH}}V(R)\left[1+ \xipbho(R)\right]\,,
\ee
where
\bea\label{eq:xibar}
    \bar{\xi}_\text{\tiny PBH}(R)
    &= \left(\frac{4}{9}\right)^2 A_s b_1^2 \bigg[ 
    \operatorname{Ci}(k_{\text{\tiny max}} R) 
-   \operatorname{Ci}(k_{\text{\tiny min}} R)
+   \frac{\cos(k_{\text{\tiny max}} R)}{k_{\text{\tiny max}}^2 R^2} 
-   \frac{\cos(k_{\text{\tiny min}} R)}{k_{\text{\tiny min}}^2 R^2} 
     \\ & \qquad\qquad\qquad
-   \frac{\sin(k_{\text{\tiny max}} R)}{k_{\text{\tiny max}}^3 R^3} (1+k_{\text{\tiny max}}^2 R^2)
+   \frac{\sin(k_{\text{\tiny min}} R)}{k_{\text{\tiny min}}^3 R^3} (1+k_{\text{\tiny min}}^2 R^2)
\bigg]\,\\
    &\approx \left(\frac{4}{9}\right)^2 A_s b_1^2 
    \left\{\begin{array}{ll}
        \ln\Delta\,, & \quad x \ll k^{-1}_{\text{\tiny{ max}}}, \\
        \ln\left(\frac{2.12}{x k_{\text{\tiny{min}}} }\right)\,, & \quad k^{-1}_{\text{\tiny{ max}}} \ll x \ll k^{-1}_{\text{\tiny{ min}}}, \\
        \mathcal{O}((x k_{\text{\tiny{min}}})^{-3})\,, & \quad x \gg k^{-1}_{\text{\tiny{ min}}}
    \end{array}\right.
\eea
and behaves thus qualitatively similarly to \eqref{eq:xiaH}.
Clustering is relevant at a scale $R$ if
\be
    \xipbho(R)\gg 1
\ee
given that the volume is at least as large as $1/\overline{n}_{\text{\tiny PBH}}$ such that it is expected to contain some PBHs, that is $\langle N(R)\rangle\gg 1$. 

\subsection{The local bias}

To fully calculate the correlation function, we need now to estimate the linear bias $b_1$. To do so, we can use the peak-background split picture in which the perturbations are divided into short- (peak) and long-wavelength (background) modes\,\cite{Sheth:1999mn}. The first is treated as a stochastic, and the second as a classical variable. 
Using the condition $\mathcal{C} > \mathcal{C}_{\text{\tiny c}}$ as a proxy for the discrete PBH distribution, the probability $P\left(\mathcal{C} > \mathcal{C}_{\text{\tiny c}}\right)$ allows us to define the non-Poisson component of the PBH fluctuation in the presence of the long mode of the Gaussian component of the curvature perturbation
$\zeta^l_\text{\tiny g}(\vec x)$:
\be
\dpbh({\vec x})=\frac{P\left(\mathcal{C} > \mathcal{C}_{\text{\tiny c}}|\zeta^l_\text{\tiny g}(\vec x)\right)}{P\left(\mathcal{C} > \mathcal{C}_{\text{\tiny c}}\right)}-1\simeq\left. \frac{\partial\ln P\left(\mathcal{C} > \mathcal{C}_{\text{\tiny c}}|\zeta^l_\text{\tiny g}(\vec x)\right)}{\partial \zeta^l_\text{\tiny g} (\vec x)}\right|_{\zeta^l_\text{\tiny g}=0}\zeta^l_\text{\tiny g} (\vec x).
\ee
On expanding Eq.\,\eqref{eq:C1expl} in terms of the short and long mode\footnote{We go beyond Ref.~\cite{vanLaak:2023ppj}, which only considered the local expansion \eqref{eq:FirstExpansion} up to cubic order and adopted peak theory. }
(where now we have made explicit the position $\vec x$ where the PBH is formed), we obtain
\begin{equation}\label{eq:C1expla}
\mathcal{C}_1(\vec x,r_m) = -\frac{4}{3}\,r_m\,\zeta_\text{\tiny g}^{s\prime}(r_m)\,
F_{\text{\tiny s}}' -\frac{4}{3}\,r_m\,\zeta_\text{\tiny g}^{s\prime}(r_m)\,F_{\text{\tiny s}}''\zeta^l_\text{\tiny g}(\vec x)=
\mathcal{C}^s_1(r_m)\left(1+\frac{F_{\text{\tiny s}}''}{F_{\text{\tiny s}}'}\zeta^l_\text{\tiny g}(\vec x)\right),
\end{equation}
with
\begin{equation}
\mathcal{C}^s_1(r_m)=F'_{\text{\tiny s}} \mathcal{C}^s_{\text{\tiny g}}(r_m)~~{\rm and}~~
\mathcal{C}^s_{\text{\tiny g}}(r_m)=-\frac{4}{3}r_m\zeta_\text{\tiny g}^{s\prime}(r_m),
\end{equation}
and the primed quantities $F_{\text{\tiny s}}'$ etc. are derivatives of $F$ evaluated at $\zeta^l_\text{\tiny g}=0$. Note that we have neglected the radial derivative of the long mode. Likewise,
\be\label{eq:C1explaa}
    \mathcal{C}(\vec x,r_m)= 
    \mathcal{C}_1(\vec x,r_m)
 - \frac{3}{8}
 \mathcal{C}_1^2(\vec x,r_m)\simeq\mathcal{C}^s_1(r_m)\left(1+\frac{F_{\text{\tiny s}}''}{F_{\text{\tiny s}}'}\zeta^l_\text{\tiny g}(\vec x)\right)-\frac{3}{8}\mathcal{C}^{s2}_1(r_m)\left(1+2\frac{F_{\text{\tiny s}}''}{F_{\text{\tiny s}}'}\zeta^l_\text{\tiny g}(\vec x)\right).
\ee
We can redefine the stochastic field
\be
    \mathcal{C}^s_{\text{\tiny g}}(r_m)
    \to \left(1+\frac{F_{\text{\tiny s}}''}{F_{\text{\tiny s}}'}\zeta^l_\text{\tiny g}(\vec x)\right)
    \mathcal{C}^s_{\text{\tiny g}}(r_m), 
\ee
in a way that the presence of the long mode is incorporated in the correlators
\begin{eqnarray}
    \sigma_{\text{\tiny c}}^2&\to &\left(1+2\frac{F_{\text{\tiny s}}''}{F_{\text{\tiny s}}'}\zeta^l_\text{\tiny g}(\vec x)\right)
    \sigma_{\text{\tiny c}}^2,\nonumber\\
    \sigma_{\text{\tiny cr}}^2&\to &\left(1+\frac{F_{\text{\tiny s}}''}{F_{\text{\tiny s}}'}\zeta^l_\text{\tiny g}(\vec x)\right) \sigma_{\text{\tiny cr}}^2,\nonumber\\
    \sigma_{\text{\tiny r}}^2&\to & \sigma_{\text{\tiny r}}^2.
\end{eqnarray}
As a result,
\be
    \dpbh({\vec x})
    \simeq\left( 2\frac{\partial\ln P(\mathcal{C} > \mathcal{C}_{\text{\tiny c}})}
    {\partial\ln \sigma_{\text{\tiny c}}^2}+\frac{\partial\ln P(\mathcal{C} > \mathcal{C}_{\text{\tiny c}})}
    {\partial\ln \sigma_{\text{\tiny cr}}^2}\right)\frac{F_{\text{\tiny s}}''}{F_{\text{\tiny s}}'}\zeta^l_\text{\tiny g} (\vec x),
\ee
where $F_{\text{\tiny s}}''/F_{\text{\tiny s}}'$ is a function of $\mathcal{C}_{\text{\tiny c}}$.
Following Ref.~\cite{Desjacques:2018wuu}, we compute the bias factor $b_1$ according to
\be
    b_1 = \frac{1}{P(\mathcal{C} > \mathcal{C}_{\text{\tiny c}})} \int_{\mathcal{D}} \frac{\left(\mathcal{C}_{\text{\tiny g}}^2 \sigma^2_{\text{\tiny r}} - \mathcal{C}_{\text{\tiny g}} \zeta_{\text{\tiny g}} \sigma^2_{\text{\tiny cr}} - \sigma^2_{\text{\tiny c}} \sigma^2_{\text{\tiny r}} + \sigma^4_{\text{\tiny cr}}\right) \exp \left(\frac{\mathcal{C}_{\text{\tiny g}}^2 \sigma^2_{\text{\tiny r}} - 2\mathcal{C}_{\text{\tiny g}} \zeta_{\text{\tiny g}} \sigma^2_{\text{\tiny cr}} + \zeta_{\text{\tiny g}}^2 \sigma_{\text{\tiny c}}^2}{2 \sigma^4_{\text{\tiny cr}} - 2 \sigma^2_{\text{\tiny c}} \sigma^2_{\text{\tiny r}}} \right)}{2 \pi \sigma_{\text{\tiny c}} \sigma_{\text{\tiny r}} \sqrt{1 - \frac{\sigma^4_{\text{\tiny cr}}}{\sigma^2_{\text{\tiny c}} \sigma^2_{\text{\tiny r}}}} \left(\sigma^2_{\text{\tiny c}} \sigma^2_{\text{\tiny r}} - \sigma^4_{\text{\tiny cr}}\right)} \frac{F_{\text{\tiny s}}''}{F_{\text{\tiny s}}'} \td\mathcal{C}_\text{\tiny g} \td\zeta_\text{\tiny g} ,
\ee
where the domain of integration $\mathcal{D}$ is defined in Eq.\,\eqref{eq:Domain}.
The bias factor can be interpreted as an average over the bivariate Gaussian distribution in the domain $\mathcal{D}$,
\begin{equation}
    b_1 = \left\langle \left(\frac{\mathcal{C}_{\text{\tiny g}}^2 \sigma^2_{\text{\tiny r}} - \mathcal{C}_{\text{\tiny g}} \zeta_{\text{\tiny g}} \sigma^2_{\text{\tiny cr}}}{\sigma^2_{\text{\tiny c}} \sigma^2_{\text{\tiny r}} - \sigma^4_{\text{\tiny cr}}} - 1\right) \left(\frac{F_{\text{\tiny s}}''}{F_{\text{\tiny s}}'}\right) \right\rangle \,.
\end{equation}
Before entering the main part of our analysis, it is instructive to pause and compare our result with existing literature.

Assuming a quadratic power series expansion for the primordial NGs, as in Eq.\,\eqref{eq:FirstExpansion}, in Ref.\,\cite{Tada:2015noa} the authors derive a bias factor linearly proportional to $f_\text{\tiny NL}$. By contrast, our approach predicts
\begin{equation}
b_1 \sim \left\langle f(\zeta_{\text{\tiny g}},\mathcal{C}_{\text{\tiny g}})\frac{f_\text{\tiny NL}}{1+\frac{6}{5}f_\text{\tiny NL}\zeta_{\text{\tiny g}}}\right\rangle.
\label{bias}
\end{equation}
This discrepancy arises because, in Ref.\,\cite{Tada:2015noa}, the authors consider the thresholded regions of the curvature perturbation field $\zeta$ to define the collapse and formation of PBHs. However, as previously discussed, the appropriate field for defining the threshold condition for collapse is the compaction function $ \mathcal{C}$. The non-linear relationship between these two quantities\footnote{The impact of such non-linearities on PBH abundance was first investigated in Refs.\,\cite{DeLuca:2019qsy, Young:2019yug}.} leads to the difference in the results.

\begin{figure}[t!]
	\begin{center}
\includegraphics[width=.49\textwidth]{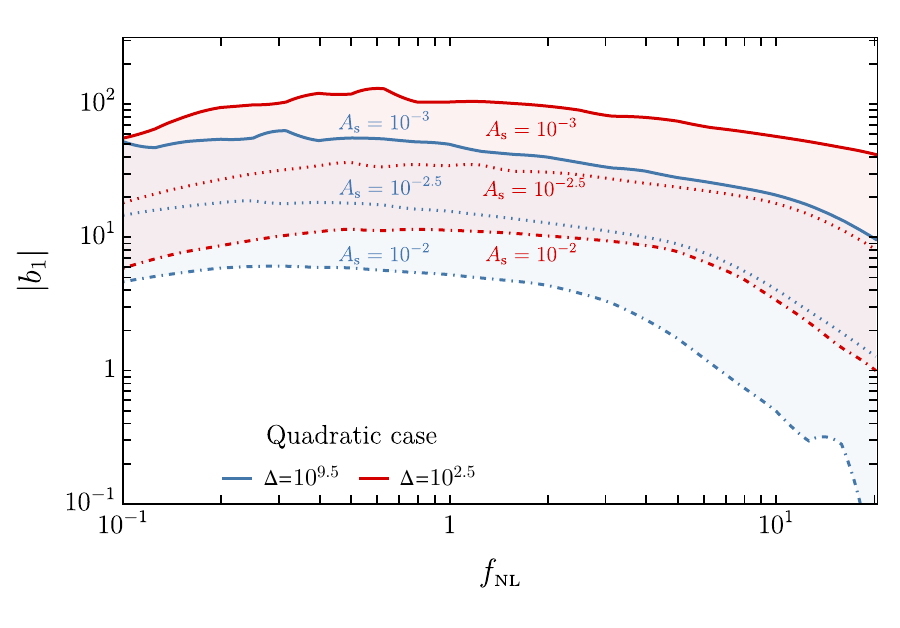}
\includegraphics[width=.49\textwidth]{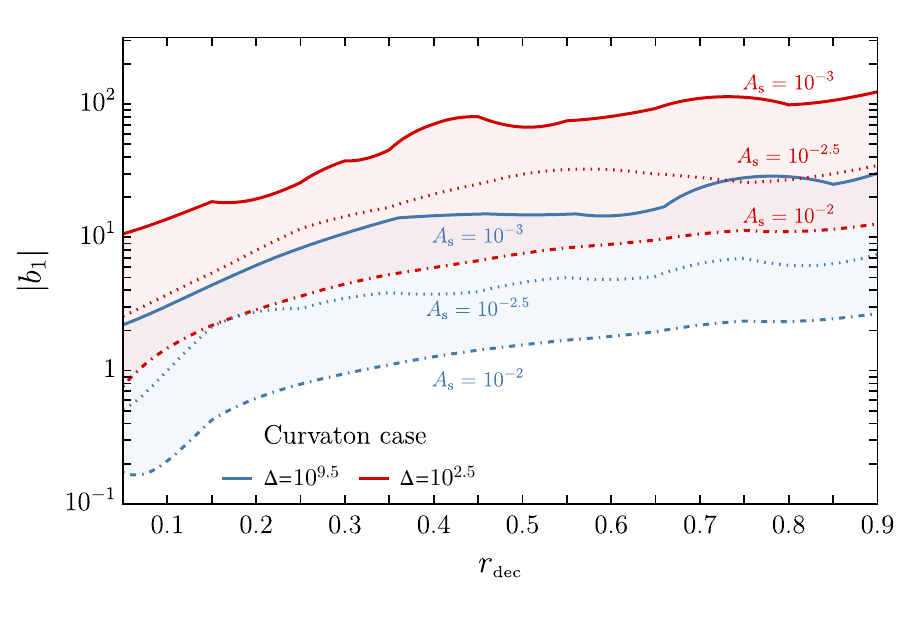}
		\caption{\em
The bias factor $b_1$ as a function of the 
NGs parameters $f_\text{\tiny NL}$ (left panel) and $r_{\rm dec}$ (right panel) for several choices of the amplitude $A$, assuming a broad (blue lines) and narrow (red lines) power spectrum.}\label{fig:biasX} 
	\end{center}
\end{figure}

In Fig.\,\ref{fig:biasX}, the bias factor $b_1$ is shown as a function of the non-Gaussian parameters $f_\text{\tiny NL}$ (left panel) and $r_{\rm dec}$ (right panel) for power spectra with two different widths.

\subsection{Some explicit scenarios}

We adopt the broad power spectrum parametrized in Eq.\,\eqref{eq:PS}, without detailing the specific inflationary model for its origin. 
To be as model-independent as possible, for primordial NGs, we consider the quadratic ansatz (see \eqref{eq:FirstExpansion}) and the curvaton scenario (see Eq. \eqref{eq:MasterX}). 
As shown in Ref.\,\cite{Ferrante:2023bgz}, in both cases, it is possible to realize a broad spectrum, which connects PBHs in the asteroid mass range to those relevant for the LVK merger events. Details about the non-Gaussian relation for the curvaton models are reported in Appendix~A. We do not consider the USR case since for the broad spectrum the primordial NGs are negligible\,\cite{Franciolini:2022pav}.
We stress that the values used for the quadratic ansatz considered in this work cover a wide range of common models for PBH production. We neglect cases with huge primordial NGs that can violate the perturbativity and are not easily realizable in literature.
These cases considered here are summarized in Table~\ref{tab:Cases}.

{
\renewcommand{\arraystretch}{1.4}
\setlength{\tabcolsep}{4pt}
\begin{table}[h]
    \centering
    \begin{tabular}{||p{12mm}|p{12mm}|p{17mm}|p{44mm}|p{17mm}|p{12mm}||}
    \hline \hline
    Cases & $\Delta $ &  $A_{\text{\tiny s}}$ & NG &  $f_\text{\tiny PBH}$ &  $b_1$ \\
       \hline \hline
       $Q_{S1}$  & $10^{2.5}$  & $\simeq 10^{-2.2}$ & Quadratic $f_\text{\tiny NL}=0.42$  &  $\simeq 5\cdot 10^{-4}$ & 18 \\
         \hline
       $Q_{S2}$  & $10^{2.5}$ & $\simeq 10^{-3.2}$ & Quadratic $f_\text{\tiny NL}=10.75$ & $\simeq 3\cdot 10^{-3}$  & 80  \\
         \hline
       $Q_{L1}$  & $10^{9.5}$ & $\simeq 10^{-2.4}$ & Quadratic $f_\text{\tiny NL}=0.42$ & $\simeq1$ & 15  \\
         \hline
       $Q_{L2}$  & $10^{9.5}$ & $\simeq 10^{-3.5}$ & Quadratic $f_\text{\tiny NL}=10.75$ & $\simeq1$ & 63 \\
         \hline
        \hline
      $C_{S1}$   & $10^{2.5}$ & $\simeq 10^{-2.0}$ & Curvaton $r_{\rm dec}=0.5$  & $\simeq 3\cdot 10^{-4}$ & 8  \\
         \hline
       $C_{S2}$  & $10^{2.5}$ & $\simeq 10^{-2.5}$  & Curvaton $r_{\rm dec}=0.1$   & $\simeq 3\cdot 10^{-4}$  & 5 \\
         \hline
      $C_{L1}$   & $10^{9.5}$ & $\simeq 10^{-2.0}$ & Curvaton $r_{\rm dec}=0.5$  & $\simeq1$ & 1.8  \\
         \hline
      $C_{L2}$   & $10^{9.5}$  & $\simeq 10^{-2.6}$ & Curvaton $r_{\rm dec}=0.1$  & $\simeq1$ & 1.5 \\
         \hline \hline
    \end{tabular}
    \caption{Cases considered in this paper. We imposed for each case $k_{\text{\tiny min}}=10^{5}$ $ {\rm Mpc}^{-1}$.}
    \label{tab:Cases}
\end{table}
}

\begin{figure}[t]
	\begin{center}
\includegraphics[width=.49\textwidth]{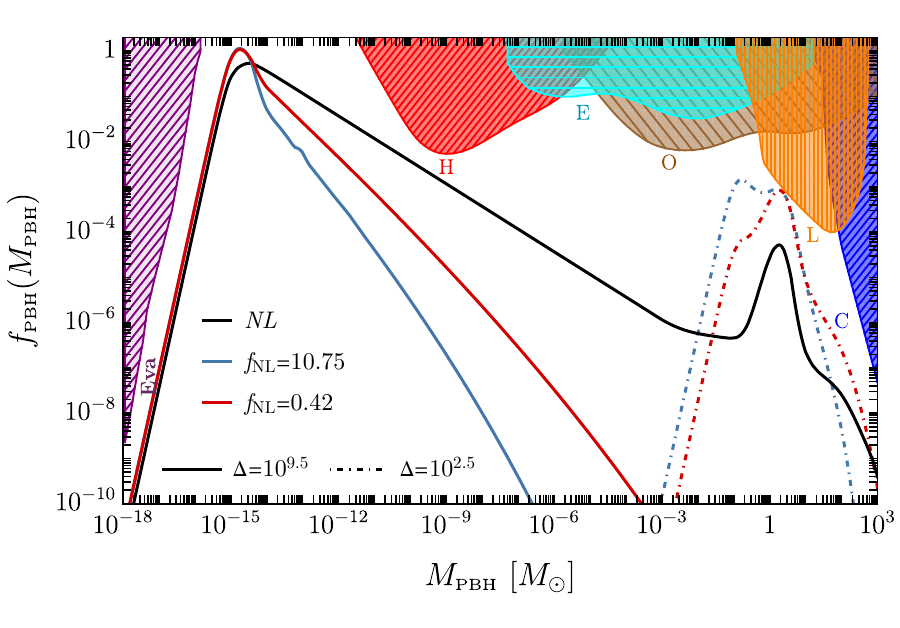}
\includegraphics[width=.49\textwidth]{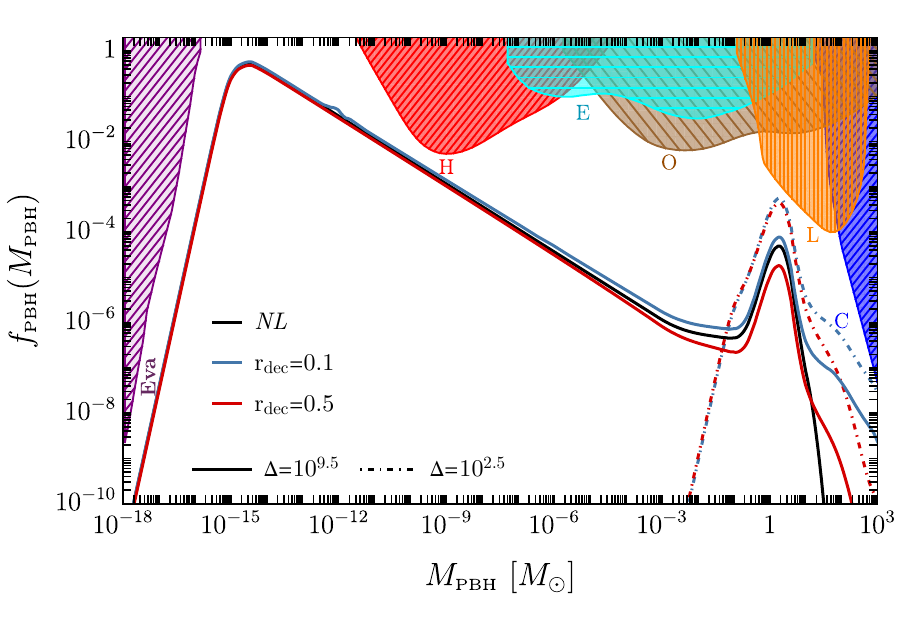}
		\caption{\em 
Two classes of mass functions resulting from power spectra with $k_{\text{\tiny min}}=10^{5}$ $ {\rm Mpc}^{-1}$ and $ \Delta =10^{9.5}$ (solid lines) or $ \Delta =10^{2.5}$ (dotdashed lines). On the left panel, we consider quadratic NGs with different values of $f_{\rm NL}$, while on the right panel, we consider the curvaton scenarios with NGs controlled by $r_{\rm dec}$. 
The amplitude of the power spectrum has been re-scaled for each value of $r_{\rm dec}$ such that PBHs do not overproduce dark matter. The only non-linear (NL) broad case (black solid line) has been obtained fixing $A_{\text{\tiny s}}=10^{-2.07}$.
The most stringent experimental constraints are shown (see description in the text).}\label{fig:AbuX} 
	\end{center}
\end{figure}

In Fig.~\ref{fig:AbuX}, the corresponding mass distributions $f_{\text{\tiny PBH}}$ for the various cases are shown. 
The following most constraining bounds are reported\footnote{Fig.~\ref{fig:AbuX} showing constraints for monochromatic PBH mass spectra, which are not directly comparable to the mass functions shown and rather provide a qualitative comparison. Following Ref.~\,\cite{Carr:2017jsz}, we checked that the constraints are not violated for the benchmark scenarios in Table.~\ref{tab:Cases}.}:
\textbf{Ev}aporation constraints (see also \cite{Saha:2021pqf,Laha:2019ssq,Ray:2021mxu}): EDGES\,\cite{Mittal:2021egv}, 
CMB\,\cite{Clark:2016nst}, INTEGRAL\,\cite{Laha:2020ivk,Berteaud:2022tws}, 511 keV\,\cite{DeRocco:2019fjq,Dasgupta:2019cae}, Voyager\,\cite{Boudaud:2018hqb}, 
EGRB\,\cite{Carr:2009jm});
microlensing constraints from the Hyper-Supreme Cam (\textbf{H}), Ref.\,\cite{Niikura:2017zjd}; 
microlensing constraints from \textbf{O}GLE, Refs.\,\cite{Niikura:2019kqi,Mroz:2024wag,Mroz:2024wia};  
constraints from modification of the CMB (\textbf{C}) spectrum due to accreting PBHs, Ref.\,\cite{Serpico:2020ehh,Agius:2024ecw};
direct constraints on PBH-PBH mergers with \textbf{L}IGO, Refs.\,\cite{Andres-Carcasona:2024wqk} (see also \cite{Raidal:2017mfl,Raidal:2018bbj,Vaskonen:2019jpv,LIGOScientific:2019kan,Kavanagh:2018ggo,Wong:2020yig,Hutsi:2020sol,DeLuca:2021wjr,Franciolini:2021tla,Franciolini:2022tfm}).

At this stage, several important observations can be made.
\begin{itemize}
    \item[{\it (i)}] As seen in Fig.,\ref{fig:biasX}, reducing the amplitude $A_s$ while keeping the NG parameter fixed increases the bias $b_1$. However, in contrast to earlier works, we find that $b_1$ is not proportional to $f_\text{\tiny NL}$, as corroborated by the theoretical discussion in the previous section. For a fixed amplitude and the same amount of quadratic NGs, the two ansatz yield identical values for $b_1$. However, since the quadratic cases require smaller power spectrum amplitudes to achieve the same abundance, this ansatz results in greater bias.
    
    \item[{\it (ii)}] Increasing the degree of NGs (i.e., decreasing the value of $r_{\rm dec}$ from 0.5 to 0.1) leads to an amplification of the mass distribution $f_{\text{\tiny PBH}}(M_{\text{\tiny PBH}})$. Consequently, to achieve the same amount of dark matter in the form of PBHs, the amplitude $A_s$ must be reduced. This trend is similar to that observed with a positive $f_\text{\tiny NL}$. Comparing the amplitudes of the power spectra in Tab.~\ref{tab:Cases}, we see that, when the full NG relation Eq.\,\eqref{eq:MasterX} is used instead of the quadratic approximation for $f_\text{\tiny NL}$ Eq.~\eqref{eq:fnlCurva}, the quadratic power series enhances PBH production. As a result, to fix the abundance, the amplitudes need to be further decreased, which is consistent with the findings of Ref.~\cite{Ferrante:2022mui}.
    
    \item[{\it (iii)}] The right panel of Fig.~\ref{fig:AbuX} demonstrates that, in the case of a broad power spectrum, the mass function exhibits two peaks: one at smaller masses associated with the maximum scale of the power spectrum, $k_{\text{\tiny max}}$, and another in the range of solar-mass PBHs.\footnote{The softening of the equation of state near the QCD transitions is expected to slightly influence the evolution of sub-horizon modes. Since this effect is mitigated by the window function, which also smooths out sub-horizon modes, it is neglected here.} When modes corresponding to these masses re-enter the cosmological horizon during the QCD phase transitions, the softening of the equation of state enhances PBH formation\,\cite{Jedamzik:1998hc, Byrnes:2018clq, Franciolini:2022tfm, Escriva:2022bwe, Musco:2023dak}. As shown in the left panel of the same figure, the secondary peak in the quadratic NG case is negligible compared to the curvaton case. This can be understood from the analysis in Ref.~\cite{Ferrante:2022mui}. For broad power spectra, in the absence of significant primordial NGs, the mass fraction $\beta$ is scale-invariant, and for a constant threshold $\mathcal{C}_{\text{\tiny c}}$, the mass distribution $f_{\text{\tiny PBH}}$ scales as $\propto M^{-3/2}_{\text{\tiny PBH}}$\,\cite{DeLuca:2020agl}. Primordial NGs break the $M_{\text{\tiny H}}$-independence of $\beta$, and the effect depends on the type and magnitude of the NGs, as shown in Fig.~\ref{fig:AbuX}. In curvaton models with large primordial NGs, PBH formation in the solar-mass range is enhanced, while the classical $\propto M^{-3/2}_{\text{\tiny PBH}}$ distribution is recovered at intermediate masses. In contrast, the quadratic ansatz tends to underproduce solar-mass PBHs relative to the non-linear case. Furthermore, for intermediate masses, the slope of the mass distribution (in modulus) increases with the amount of quadratic NGs.
\end{itemize}

\section{Clustering and merger length scales}

Suppose that PBHs are born clustered inside a sphere of radius $\sim k^{-1}_{\text{\tiny min}}$. 
In radiation domination, the cluster does not evolve to form a halo  as long as \cite{Kolb:1994fi,DeLuca:2022bjs}
\be
    f_{\text{\tiny PBH}}\,\bar\xi_{\text{\tiny PBH}}(k^{-1}_{\text{\tiny min}})\ltap 1.
\ee
This is because, in the mass range under consideration, the $f_{\text{\tiny PBH}}$-dependent PBH abundance is too small relative to the dark matter density, and the overdensity of the cluster is insufficient to overcome the radiation pressure and produce a bound matter-dominated region following gravitational collapse \cite{Kolb:1994fi}.

Once the matter-dominated period begins, PBH perturbations decouple from the Hubble flow, collapse, and virialize to form halos with a virial density about 200 times the background density at the time of virialization. The formation of PBH halos is hierarchical like in a standard CDM cosmology:
the small mass PBH halos that form first are the progenitors of the more massive PBH halos that virialize at later times. 
If clustering is relevant up to a given scale $\sim k^{-1}_{\text{\tiny min}}$, one expects that a typical PBH halo will contain an average number of PBH equal to $\langle N\rangle\sim \bar n_{\text{\tiny PBH}}V(k_{\text{\tiny min}})(1+\bar\xi_{\text{\tiny PBH}}(k^{-1}_{\text{\tiny min}}))$.

During the radiation era, PBH binaries form along with the first PBH clusters. 
These binaries are likely to be highly eccentric since they acquire their angular momentum through tidal torques exerted by surrounding inhomogeneities. 
In clustered PBH scenarios, these early binaries are more prone to collisions with neighbouring PBHs or to absorption in the early clusters that tend to have frequent collisions between the PBHs. 
Even mild disturbances can extend the coalescence times of eccentric PBH binaries by several orders of magnitude, which strongly suppresses their merger rate. 
When this occurs, the PBH merger rate is driven by the formation of PBH binaries via the early three-body channel. This produces less eccentric binaries and yields merger rates that are less affected by binary-BH interactions~\cite{Vaskonen:2019jpv, Raidal:2024bmm}. As a result, the population of PBH binaries as well as their merger rate would be quite different from that expected in the absence of initial clustering, which is often assumed in the literature.

To gauge whether clustering impacts the merger of two PBHs, let us consider the typical scales involved in a merger. It is straightforward to verify that the gravitational interaction between two nearby, isolated PBHs drives their dynamical evolution if their average mass exceeds the amount of matter enclosed in a comoving sphere of radius equal to their separation. This situation can arise in the radiation-dominated era owing to the different time dependencies of the two competing effects that influence the PBHs' separation: their mutual gravitational attraction and the pull of cosmic expansion. More precisely, assuming that all PBHs have the same mass $M$, two neighbouring PBHs will detach from the Hubble flow during the radiation-dominated era if their comoving separation satisfies\,\cite{Sasaki:2016jop, Ali-Haimoud:2017rtz, Raidal:2018bbj}
\be\label{eq:Rmax}
    R < R_{\text{\tiny max}}\ltap\left(f_{\text{\tiny PBH}}/\overline{n}_{\text{\tiny PBH}}\right)^{1/3}\simeq 0.31 \left(\frac{M}{\Msun}\right)^{1/3} {\rm kpc}.
\ee
As the radius of the Hubble patch under consideration at the time of binary formation is much bigger than $1/k_{\text{\tiny min}}$, the contribution of the clustering to the average PBH number density within that patch is negligible. Therefore, the above condition also holds when clustering is present~\cite{Atal:2020igj}.

In addition, there exists a minimum separation, $R_{\text{\tiny min}}$, below which a binary with
any orbital parameter would have already merged. This corresponds to PBH configurations that result in circular orbits. Specifically, considering the current merger rate (\( t_0 \approx 14 \, \text{Gyr} \)), we find 
\begin{equation}\label{eq:Rmin}
    R_{\text{\tiny min}} \sim 9.5 \cdot 10^{-3} \left(\frac{M}{\Msun}\right)^{7/16} {\rm kpc}. 
\end{equation}
Clustering is therefore relevant in volumes of radius $R$ if the condition
\be
    \bar{\xi}_{\text{\tiny PBH}}(R)\gg 1\,\,\, {\rm for}\,\,\, R_{\text{\tiny min}}\lesssim R\lesssim R_{\text{\tiny max}}
\ee
is satisfied. We remark that the magnitude of the effect on binary formation depends on the formation channel. For instance, in the (approximately) Poisson scenarios, the most relevant scale for the 2-body binary formation channel is about $\mathcal{O}(10 R_{\rm min})$ due to the high initial eccentricity ($j \equiv \sqrt{1-e^2} = \mathcal{O}(10^{-2})$) of these binaries~\cite{Raidal:2018bbj, Raidal:2024bmm}.

\begin{figure}[t]
	\begin{center}
\includegraphics[width=.8\textwidth]{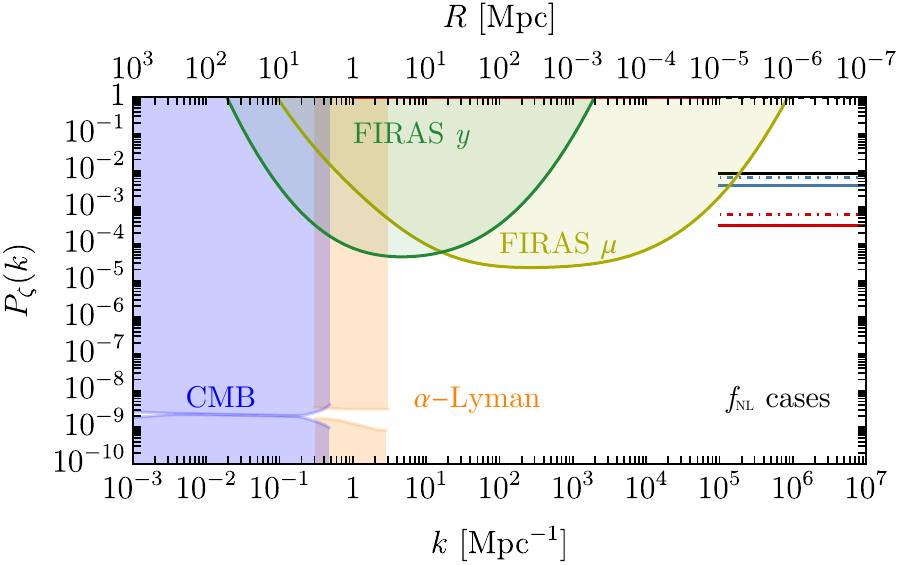}
		\caption{\em Constraints on the amplitude of the power spectrum, assuming negligible primordial NGs, from the FIRAS experiments\,\cite{Fixsen:1996nj,Chluba:2012gq,Chluba:2012we,Chluba:2013dna,Bianchini:2022dqh,Iovino:2024tyg,Wang:2025hbw}, CMB\,\cite{Planck:2018vyg}, $\alpha$-Lyman\,\cite{lyman} for a log-normal power spectrum with width $\sigma=1$ that mimics the broad power spectra assumed in this work. For clarity the $f_{\text{\tiny NL}}$ cases reported in Tab.\ref{tab:Cases} are shown.}
        \label{fig:Firas} 
	\end{center}
\end{figure}

The main point of the paper now comes into focus. To set the maximum scale $\sim k_{\text{\tiny min}}^{-1}$ below which clustering is not negligible, we must carefully examine the CMB constraints on the power spectrum of the curvature perturbation. 
As a matter of fact, CMB observations strongly constrain $P_\zeta(k)$ at scales $10^{-4} \,\mathrm{Mpc}^{-1} \lesssim k \lesssim 1\, \mathrm{Mpc}^{-1}$\,\cite{Planck:2018vyg,lyman}. 
At smaller scales other constraints must be applied,  coming from the fact that at redshifts $z \lesssim 10^6$ energy injections into the primordial plasma cause persisting spectral distortions in the CMB \cite{Fixsen:1996nj,Chluba:2012gq,Chluba:2012we,Chluba:2013dna,Bianchini:2022dqh}. These distortions are divided into chemical potential $\mu$-type distortions created higher $z$ and Compton $y$-type distortions created at $z \lesssim 5 \times 10^4$.
For a given curvature power spectrum $\mathcal{P}_{\zeta}(k)$ the spectral distortions are~\cite{Chluba:2012we,Chluba:2013dna}
\be
    X=\int_{k_{\rm m}}^{\infty} \frac{\mathrm{d} k}{k} \mathcal{P}_{\zeta}(k) W_X(k)\,,
\ee
with $X=\mu, y$, while $k_{\rm m}=1$ Mpc$^{-1}$ and the window functions\footnote{Including NGs modifies only the tails of the constraints and not the most constrained region~\cite{Sharma:2024img}.} can be approximated by
\be
    W_\mu(k)
    \simeq 2.2\left[e^{-\frac{(\hat{k} / 1360)^2}{1+(\hat{k} / 260)^{0.6}+\hat{k} / 340}}-e^{-(\hat{k} / 32)^2}\right], 
    \qquad
    W_y(k)
    \simeq 0.4 \,e^{-(\hat{k} / 32)^2}\,,
\ee
with $\hat{k}=k /\left( \mathrm{Mpc}^{-1}\right)$.
The COBE FIRAS measurements constrain the $\mu$ distortions such as $\mu \leq 4.7 \times 10^{-5}$~\cite{Bianchini:2022dqh} and $y\leq 1.5 \times 10^{-5}$ at the $95\%$ confidence level~\cite{Fixsen:1996nj}.

As shown in Fig.\,\ref{fig:Firas}, requiring sufficiently large curvature perturbations, i.e. $A_{\text{\tiny s}}\gtrsim10^{-3}$, to obtain a sizeable abundance of PBHs at the relevant scales, the FIRAS constraints set the limit~\cite{Iovino:2024tyg}
\be
    k^{-1}_{\text{\tiny min}}\ltap R_{\text{\tiny FIRAS}}\simeq  10^{-2}\,{\rm kpc}
\ee
on broad spectra, which is already a factor ${\cal O}(10)$ smaller than $R_{\text{\tiny max}}$. 
The question, then, is: can $\bar{\xi}_{\text{\tiny PBH}}(R)$ be much larger than unity in the range $R_{\text{\tiny min}}\ltap R\ltap 10^{-2} {\rm kpc} \simeq R_{\text{\tiny FIRAS}}$?

\begin{figure}[t]
	\begin{center}
\includegraphics[width=.49\textwidth]{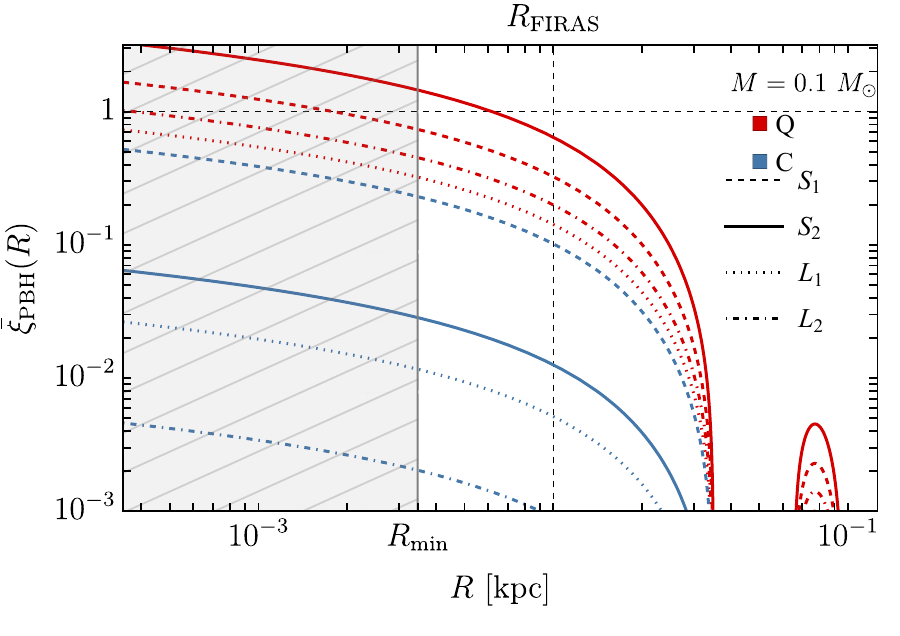}
\includegraphics[width=.49\textwidth]{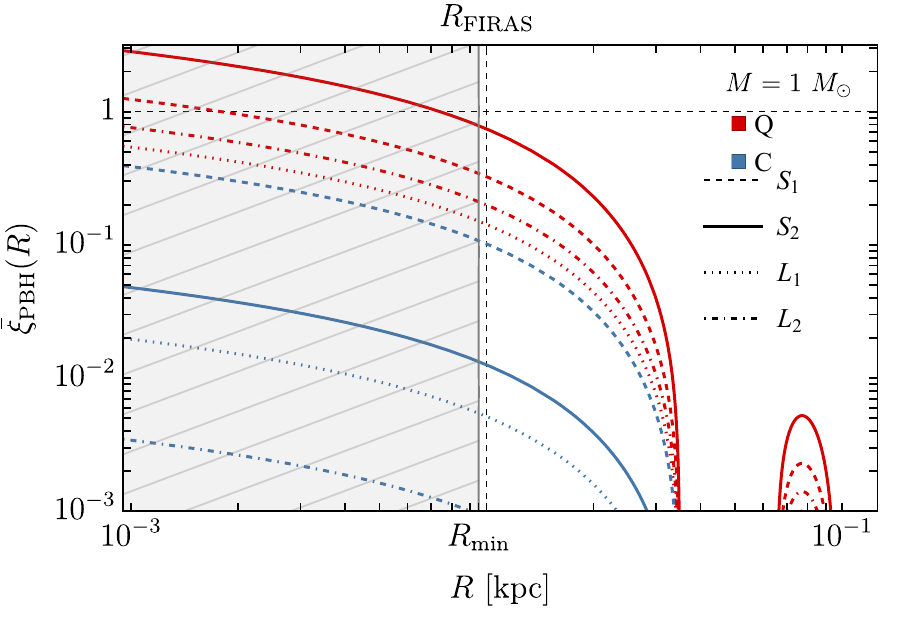}
		\caption{\em 
The correlation length $\bar{\xi}_{\text{\tiny PBH}}(R)$ in the relevant range of scales for the benchmark cases reported in Tab.\,\ref{tab:Cases}. We focus on the masses for the binary, respectively $M=0.1\,\Msun$ (left panel) and $M=1\,\Msun$ (right panel).
The shaded region indicates scales which do not affect binaries merging today. This figure assumes $k_{\text{\tiny min}}^{-1} = R_{\text{\tiny FIRAS}} $. }\label{fig:xiR} 
	\end{center}
\end{figure}

In Fig.\,\ref{fig:xiR}, we show how the required $\bar{\xi}_{\text{\tiny PBH}}(R)$ changes with $R$ for the different benchmark cases reported in Tab.\,\ref{tab:Cases} and for two masses, $M=0.1\,\Msun$ and $M=1\,\Msun$. For PBHs heavier than the Sun, we have $R_{\text{\tiny FIRAS}}<R_{\rm min}$, making the argument even stronger.
As it is clear from this figure, in the relevant range of scales and for masses of interest for LVK\,\cite{LIGOScientific:2018mvr, LIGOScientific:2020ibl, KAGRA:2021vkt}, i.e. $M\gtrsim0.1\,\Msun$, the initial spatial clustering is irrelevant, given the strong constraints on the PBH abundance in such mass range and the fact that the merger depends on the combination $f_{\text{\tiny PBH}}(1+\bar{\xi}_{\text{\tiny PBH}})$~\cite{Raidal:2024bmm}.

However, if one considers SGWBs induced by PBH mergers and galactic PBH binaries, LVK O5 and ET can in principle be sensitive to lower masses and probe PBH scenarios involving PBHs as light as $10^{-2}\Msun$ and $10^{-6}\Msun$, respectively~\cite{Pujolas:2021yaw} (see also \cite{vanDie:2024htf} for LISA). We find that for $M = 10^{-6}\Msun$, where $f_{\text{\tiny PBH}}$ may acquire values as large as about 0.1, the volume average correlation is only bounded by $\bar{\xi}_{\text{\tiny PBH}}(R) \lesssim \mathcal{O}(10)$, which suggests that clustering might play a role for lighter PBHs accessible by future GW experiments. In particular, as non-negligible correlations can be present at scales relevant for binary formation, PBH binary formation and the resulting merger rate may be enhanced. If, furthermore, the correlations are less prominent at scales associated with the formation of $N \gg 2$ PBH clusters, the subsequent clustering evolution will approximately follow the usual Poisson case, and so will also the disruption of PBH binaries within clusters formed during matter domination.

\section{Conclusions}

Discovering the presence of PBHs is one of the main targets of the new era of gravitational wave astronomy. Assessing the merger rate of PBH binaries has become therefore imperative. In this paper, we have made a simple, but relevant remark, that the initial clustering of PBH in the mass range relevant for LVK does not affect the standard estimate of the PBH binary merger rate~\cite{Raidal:2024bmm} which assumes an initial PBH Poisson distribution. Our conclusions apply to common single-field models in which PBHs are formed via the collapse of sizeable fluctuations generated during inflation. They are also robust in the sense that we have assumed a broad power spectrum for the curvature perturbation, thus maximally enhancing the possible clustering.

Here we stress that our claim does not apply for multi-field inflation scenarios\,\cite{Tada:2015noa,Atal:2019cdz,Suyama:2019cst}, where the curvature perturbation can acquire a large-scale modulation from a field that is not the primary source of the perturbation. Moreover, PBH may also form in alternative scenarios~\cite{Flores:2024eyy}. 
For the case of phase transitions (PTs)~\cite{Lewicki:2023ioy, Lewicki:2023mik, Lewicki:2024ghw, Lewicki:2024sfw}, for example, the correlation length is limited by causality as PTs cannot produce super-Hubble correlations at formation time. The relevant scales that can significantly affect PBH binaries are much larger. So, we do not expect this scenario to avoid our conclusion. 
We leave the investigation of alternative cases for future work.

In this paper, we have focused on PBH masses reachable by the LVK collaboration. As it is clear from the mass scalings, our no-go argument does not apply to lighter PBH masses, whose binaries could lead to merger rates potentially detectable by ET\,\cite{Branchesi:2023mws, Franciolini:2023opt} by ultra-high frequency detector\,\cite{Pujolas:2021yaw,Franciolini:2022htd,Aggarwal:2025noe}. We leave this analysis for future work.

\section*{Acknowledgements}
A.J.I., F.C, and G.P. thank the University of Geneva for the kind hospitality during the realization of this project. F.C. acknowledges the financial support provided under the ”Progetti per Avvio alla Ricerca Tipo 1”, protocol number AR12419073C0A82B.
A.R. acknowledges support from the Swiss National
Science Foundation (project number CRSII5 213497) and from the Boninchi Foundation for the
project “PBHs in the Era of GW Astronomy”. 

\appendix
\setcounter{equation}{0}
\setcounter{section}{0}
\setcounter{table}{0}
\makeatletter
\renewcommand{\theequation}{A\arabic{equation}}
\section*{Appendix A: Primordial Non-Gaussianities in explicit models}\label{app:Curvaton}
When presenting results inspired by the curvaton model, we will focus on primordial NG (derived analytically within the sudden-decay approximation~\cite{Sasaki:2006kq})
\be\label{eq:MasterX}
    \zeta = \ln\big[X(r_{\rm dec},\zeta_\text{\tiny g})\big]\,,
\ee
with
\begin{subequations}
\begin{align}\label{eq:XFunction}
    X &\equiv \frac{\sqrt{K}\left(1 + \sqrt{A K^{-\frac32}-1}\right)}{(3+r_{\rm dec})^{\frac13}}, 
    \\
    K & \equiv \frac{1}{2}\left((3+r_{\rm dec})^{\frac13}(r_{\rm dec}-1)P^{-\frac13} + P^{\frac13}\right), 
    \\
    P &\equiv A^2 + \sqrt{A^4 + (3+r_{\rm dec})(1-r_{\rm dec})^3}\,,
    \\
    A &\equiv \left(1 + \frac{3\zeta_\text{\tiny g}}{2r_{\rm dec}} \right)^{2}r_{\rm dec}\,.
\end{align}
\end{subequations}
The parameter $r_{\rm dec}$ is the weighted fraction of the curvaton energy density $\rho_{\phi}$ to the total energy density at the time of curvaton decay, defined by
\be
    r_{\rm dec} \equiv 
    \left.\frac{3 \rho_{\phi}}{3 \rho_{\phi} + 4 \rho_{\gamma}}\right|_{\rm curvaton\,\,decay}\,,
\ee
where $\rho_{\gamma}$ is the energy density stored in radiation after reheating. 
Thus, $r_{\rm dec}$ depends on the physical assumptions about the physics of the curvaton within a given model.
If we expand Eq. \eqref{eq:MasterX} to the second order in $\zeta_\text{\tiny g}$ we find \cite{Bartolo:2003jx}

\begin{equation}\label{eq:fnlCurva}
f_\text{\tiny NL}=\frac{5}{3}\left(\frac{3}{4 r_{\rm dec}} - 1 - \frac{r_{\rm dec}}{2}\right).
\end{equation}

\bibliographystyle{JHEP}
\bibliography{Draft}

\end{document}